\documentclass[runningheads]{llncs}
\usepackage[T1]{fontenc}
\usepackage{graphicx}
\usepackage{booktabs}

\usepackage{amsmath}
\usepackage[colorlinks,
            linkcolor=blue,
            anchorcolor=blue,
            urlcolor = blue,
            citecolor=blue
            ]{hyperref}

\usepackage[misc]{ifsym} 

\begin{document}
\title{RoCoSDF: Row-Column Scanned Neural Signed Distance Fields for Freehand 3D Ultrasound
Imaging Shape Reconstruction }
\titlerunning{RoCoSDF: Freehand 3D Ultrasound Neural Shape Reconstruction}
%
\author{Hongbo Chen\inst{1} \and Yuchong Gao\inst{1} \and Shuhang Zhang\inst{1} \and Jiangjie Wu\inst{1} \and 
\\ Yuexin Ma\inst{1} \and Rui Zheng\inst{1,2}$^{(\textrm{\Letter})}$}

\authorrunning{H. Chen et al.}

%
\institute{School of Information Science and Technology, ShanghaiTech University, \\ Shanghai, China \\
\email{zhengrui@shanghaitech.edu.cn} \and
Shanghai Engineering Research Center of Intelligent Vision and Imaging, ShanghaiTech University, Shanghai, China}

%
\maketitle              
\begin{abstract}

The reconstruction of high-quality shape geometry is crucial for developing freehand 3D ultrasound imaging. 
However, the shape reconstruction of multi-view ultrasound data remains challenging due to the elevation distortion caused by thick transducer probes.
In this paper, we present a novel learning-based framework RoCoSDF, which can effectively generate an implicit surface through continuous shape representations derived from row-column scanned datasets. 
In RoCoSDF, we encode the datasets from different views into the corresponding neural signed distance function (SDF) and then operate all SDFs in a normalized 3D space to restore the actual surface contour.
Without requiring pre-training on large-scale ground truth shapes, our approach can synthesize a smooth and continuous signed distance field from multi-view SDFs to implicitly represent the actual geometry.
Furthermore, two regularizers are introduced to facilitate shape refinement by constraining the SDF near the surface.
The experiments on twelve shapes data acquired by two ultrasound transducer probes validate that RoCoSDF can effectively reconstruct accurate geometric shapes from multi-view ultrasound data, which outperforms current reconstruction methods.
Code is available at \url{https://github.com/chenhbo/RoCoSDF}.

\end{abstract}
\section{Introduction}
\begin{figure}[tp]
	\centering
	\includegraphics[width=11cm]{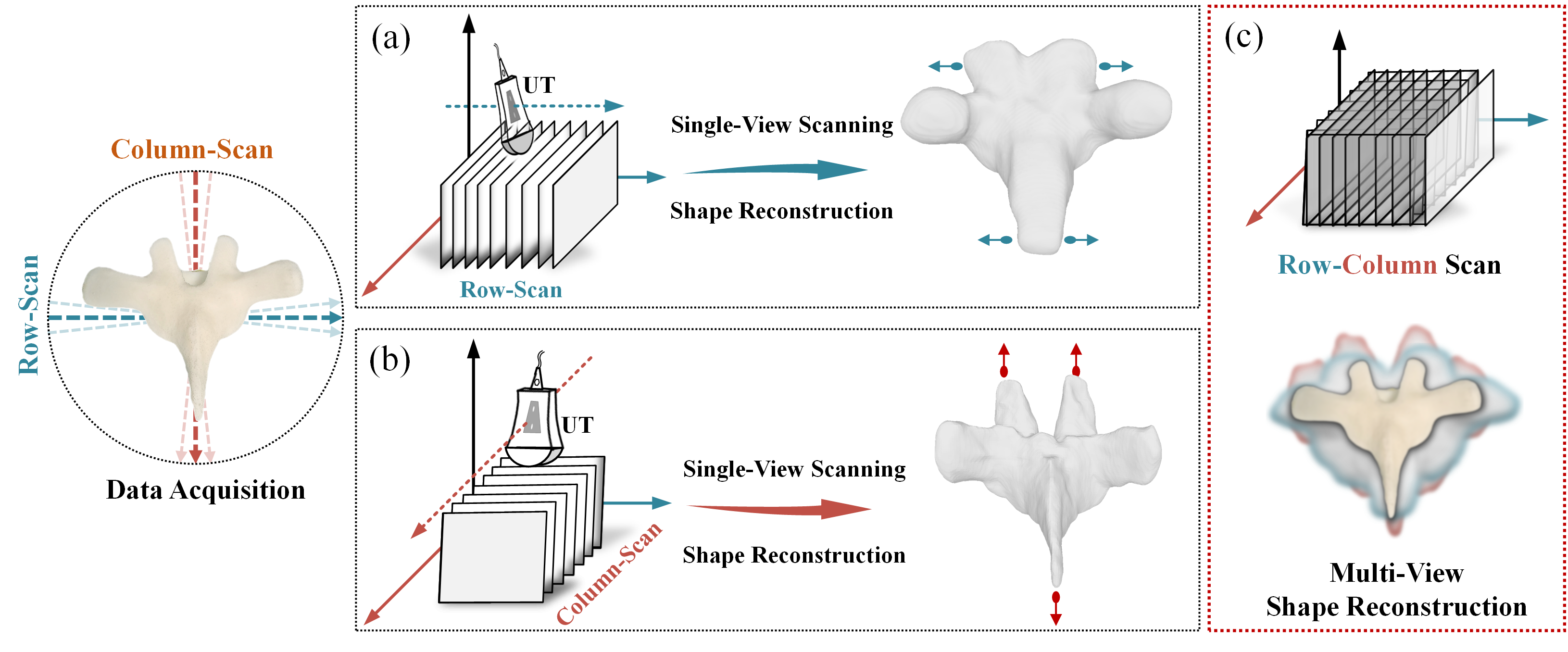}
	\caption{Two typical data acquisition manners for freehand 3D US imaging using one ultrasound transducer (UT). 
 (a) and (b) Single-view scanning and shape reconstruction in row and column directions from UNSR~\cite{chenNeuralImplicitSurface2024}. 
 (c) Proposed row-column scan and multi-view shape reconstruction.
 The shape contour (dark) is optimized from row-column scan.
 In practical use for handheld scans, our row-column scan is not necessarily orthogonal, as indicated by the dashed blue lines and red lines in the left circle.}
	\label{Fig_RoCoScan}
\end{figure}

Freehand 3D Ultrasound (US) imaging has gained considerable attention in clinical diagnostics due to its  
flexibility,  portability and large field-of-view imaging~\cite{hennerspergerComputationalSonography2015a,mohamedSurvey3DUltrasound2019,luoRecONOnlineLearning2023a,luoMultiIMUOnlineSelfconsistency2023}.
Most current freehand imaging systems usually collect 2D images and 3D poses
by scanning the object in a single-view direction (Fig.~\ref{Fig_RoCoScan}), 
utilizing a tracking device attached to the ultrasound transducer (UT)~\cite{mohamedSurvey3DUltrasound2019}.
Accurate reconstruction of geometric shape or surface is essential for freehand 3D US imaging in clinical settings,
particularly for US hard tissue imaging because the US signal is unable
to penetrate the boundaries of hard tissues~\cite{huangAnatomicalPriorBased2022,chenNeuralImplicitSurface2024}.
However, single-view scanning provides restricted perspectives, often resulting in insufficient anatomical information and distorted structures in the scanning direction, as illustrated in Fig.~\ref{Fig_RoCoScan}.

Multi-view US scanning can integrate multiple viewpoints, offering comprehensive spatial understanding and complex anatomical shapes.
However, the view-dependent nature of US imaging introduces additional challenges for 3D reconstruction, since imaging a target from multiple scanning directions can produce different reflected intensities.
Current existing multi-view reconstruction methods 
mainly focus on two strategies for freehand or robotic scanning:
 orientation-based intensity compounding ~\cite{parkImprovingThreedimensionalAutomated2023,wrightFastFetalHead2023,hennerspergerComputationalSonography2015a} 
and neural radiance field (NeRF) based view synthesis~\cite{wysockiUltraNeRFNeuralRadiance2023,gaitsUltrasoundVolumeReconstruction2024}.
Although orientation-based methods can partially address the view-dependent issue,
their results are limited by the spatial resolution and discrete pixel connectivity problem~\cite{mohamedSurvey3DUltrasound2019}. 
On the other hand, the neural radiance field (NeRF)-based method, as an emerging technology, offers the advantage of being resolution-independent in representation. However, it struggles to accurately capture the geometric shapes~\cite{zhaEndoSurfNeuralSurface2023,batlleLightNeuSNeuralSurface2023}.
Furthermore, the development of these methods in freehand scanning is more challenging because slight motion interference may lead to mismatches in multi-view fusion.

With the advance of neural implicit function, 
3D continuous shape representation has recently emerged as a useful technique for improving the geometric appearance in computer vision and medical imaging~\cite{parkDeepSDFLearningContinuous2019,wang2021neus,amiranashviliLearningContinuousShape2024a,wiesnerGenerativeModelingLiving2024}.
These representations are commonly parameterized as a neural network that
map 3D coordinates to implicit values, such as signed distance function (SDF)~\cite{parkDeepSDFLearningContinuous2019,NeuralPull,marschnerConstructiveSolidGeometry2023}. 
Inspired by~\cite{NeuralPull}, a self-supervised learning method has recently been reported for freehand 3D US neural surface reconstruction (UNSR)~\cite{chenNeuralImplicitSurface2024}.
It trains a Multi-Layer Perceptrons (MLPs) network to  
learn neural SDF from input US volumetric masks without the requirements of ground truth signed distance fields, point cloud normals, and occupancy fields. 
However, the limitations of single-view imaging hinder UNSR from learning precise structure.
Fig.~\ref{Fig_RoCoScan}(a) and (b) illustrate that the overall vertebra shape is elongated in the 
row and column direction by such a method.

To address these limitations, we propose RoCoSDF: \textbf{Ro}w-\textbf{Co}lumn scanned neural \textbf{S}igned \textbf{D}istance \textbf{F}unction fields,  
a novel framework based on neural implicit functions for shape reconstruction of multi-view freehand 3D ultrasound imaging.
As shown in Fig.~\ref{Fig_RoCoScan}(c), RoCoSDF aims to synthesize an SDF field to implicitly represent the actual structures without ground truth shape supervision by utilizing two typical scan views in row and column directions.
Specifically, given the row-column segmented points from a shape, 
we first learn two SDFs for the row-scan and column-scan
using a self-supervised strategy in a normalized space.
Then, constructive solid geometry (CSG) is employed to extract a signed distance field from the two SDFs, which serves as the initial representation to be refined further.
We subsequently sample the query points and SDF values in the SDF field for shape refinement using a supervised strategy. 
Particularly, 
we introduce two regularizers to enhance the learning of a more accurate neural SDF field.
Our proposed method is quantitatively and qualitatively validated on twelve vertebrae data scanned by two UTs. 
The results demonstrate superior shape fidelity over existing approaches.

\section{Methodology}
\subsection{Overview}
\begin{figure}[tp]
	\centering
	\includegraphics[width=10cm]{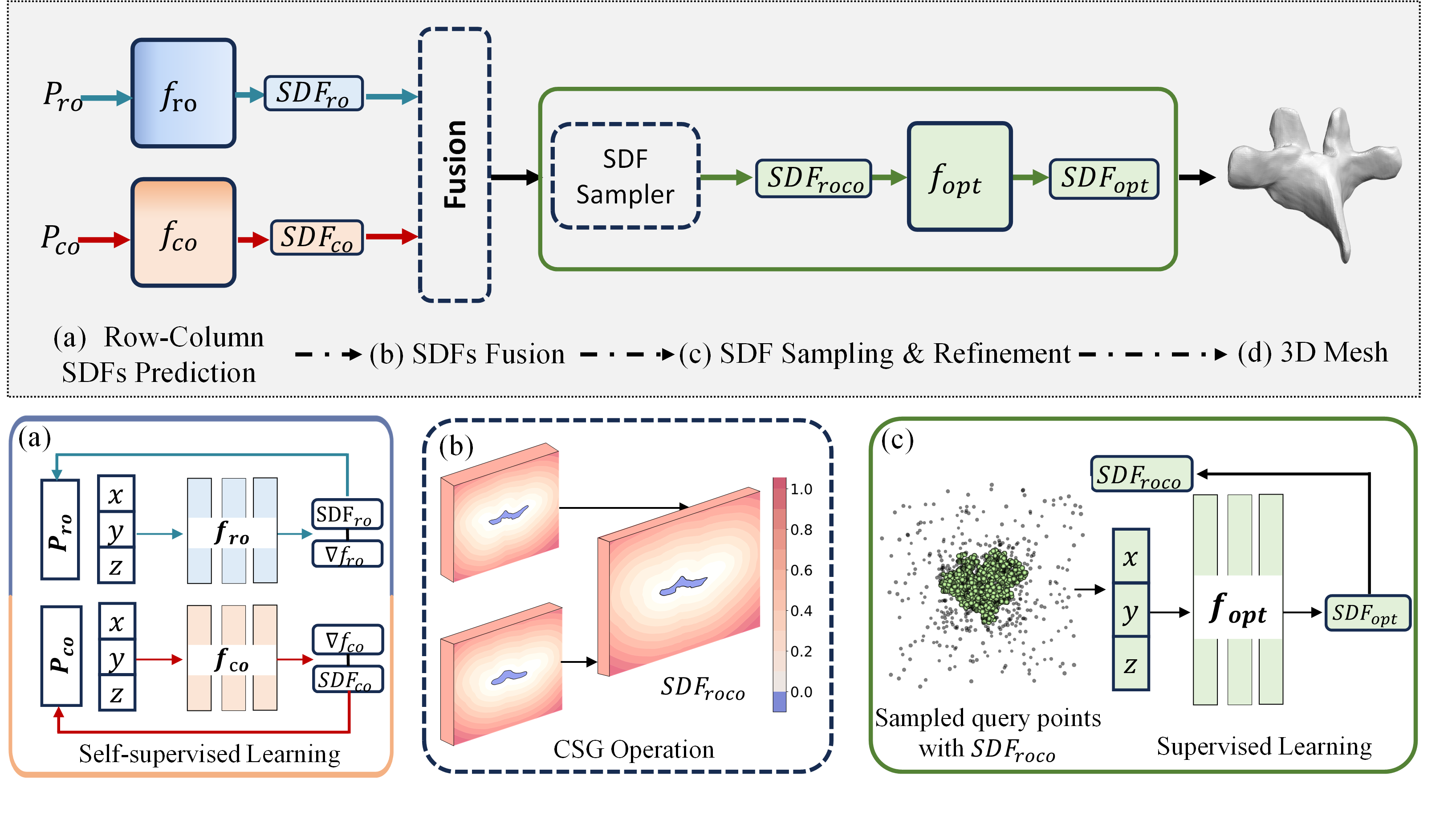}
	\caption{An overview of the proposed framework for shape reconstruction.  
        (a) Row-Column neural SDFs prediction from  point cloud $P_{ro}$ and $P_{co}$. 
         (b) SDFs fusion using constructive solid geometry (CSG). 
         (c) SDF sampling and refinement.
         }
	\label{Fig_RoCoSDF}
\end{figure}

An overview of the proposed framework for multi-view neural shape reconstruction from row-column scan
is illustrated in Fig.~\ref{Fig_RoCoSDF}.
Our coarse-to-fine reconstruction pipeline primarily comprises four steps.
In Step (a), separate training is conducted to predict the row and column SDFs from input row-column datasets. 
Step (b) aims to fuse the predicted  SDFs to obtain a signed distance field,
and step (c) involves sampling the query points and SDF value in the distance field
for shape refinement.
Finally, step (d) extracts the 3D mesh from the optimized SDF using the Marching Cube algorithm~\cite{lorensenMarchingCubesHigh1987a}.

\subsection{RoCoSDF: Multi-view Neural Shape Reconstruction}

\subsubsection{Row-Column Neural SDFs Prediction}

A neural network $f_\theta$, such as MLP, can be trained as an implicit function to map any 3D point coordinate, $\mathbf{x} = [x, y, z] \in R^3$, to its corresponding SDF value of $s \in R$.
Here, 
$\theta$ represents the learnable parameters of the network.
The object surface  $\mathcal{S}$ is implicitly represented by the zero-level-set of neural SDFs,  $f_\theta(\cdot)$ = 0.
Each 3D shape can be individually parameterized by a $f_\theta(\cdot)$.

\begin{equation}
\begin{aligned}
& f_\theta(\mathbf{x})=s, \left\{\begin{aligned}
d(\mathbf{x}, \mathcal{S}), & \text { if } \mathbf{x} \text { is outside of } \mathcal{S} \\
0, & \text { if } \mathbf{x} \text { is on the } \mathcal{S}\\
-d(\mathbf{x}, \mathcal{S}),  & \text { if } \mathbf{x} \text { is inside of } \mathcal{S}
\end{aligned}\right. 
\end{aligned}
\end{equation}
 where $d(\mathbf{x}, \mathcal{S})$ is the positive 
 distance from $\mathbf{x}$ to surface $\mathcal{S}$.
The sign before $d(\mathbf{x}, \mathcal{S})$ is positive (negative) when $\mathbf{x}$ reaches the surface from outside (inside) of object.

A query point $\mathbf{x}$ can be projected to its nearest point $\mathbf{x^{\prime}}$ on the surface $\mathcal{S}$ along or against the network 
gradient ($\nabla f_\theta(\mathbf{x})$) according to the predicted signed distance.
\begin{equation}\label{EqProjection}
	{\mathbf{x}}^{\prime}=\mathbf{x} - f_\theta(\mathbf{x}) 
  \cdot \nabla f_\theta(\mathbf{x})
\end{equation}

Here, we utilize two MLPs neural networks
to predict the row-column neural SDFs for row-scan and column-scan, respectively.

\begin{equation}
\begin{aligned}
& \operatorname{SDF}_{\text {ro }}(\mathbf{x})=f_{ro}(\mathbf{x}) \\
& \operatorname{SDF}_{\text {co }}(\mathbf{x})=f_{co}(\mathbf{x})
\end{aligned}
\end{equation}
where $f_{ro}$ and $f_{co}$ are the row-scan SDF decoder and the column-scan SDF decoder, respectively

In step (a), the $SDF_{ro}$ and $SDF_{co}$ map 
 the 3D query points generated from dual-view point clouds
 $P_{ro}$ and $P_{co}$ to a normalized signed distance field using MLPs
by a series of self-supervised loss functions with regularizers.

\subsubsection{Signed Distance Fields Fusion }
Constructive Solid Geometry (CSG) commonly includes three
boolean operations: union, intersection and difference, which are computing on the implicit functions~\cite{sharmaCSGNetNeuralShape2018,marschnerConstructiveSolidGeometry2023}.
With row and column signed distance fields, an SDF can be coarsely fused through CSG operation
by operating the intersection between $SDF_{ro}$ and $SDF_{co}$.
\begin{equation}
\text{Intersection} \ f_{ro} \cap f_{co}: SDF_{roco} =\max \left(SDF_{ro}, SDF_{co}\right)
\end{equation}

After the fusion, the actual object shape can be naively 
reconstructed using $SDF_{roco}$. 
However, the direct CSG operation on row-column SDFs will lead to 
a rough local surface and unexpected errors since CSG is commonly designed for primitive shape elements.

\subsubsection{SDF Sampling and Refinement}
To refine $SDF_{roco}$, 
we further train a $f_{opt}$ to optimize the fused distance fields in a supervised way.
An SDF sampler is designed to 
directly sample the query points and SDF values 
in a 3D cube of $SDF_{roco}$.
We randomly generate the query points within the cube space ranging from -1 to 1.
Each query point is fed into the $SDF_{roco}$ to obtain the corresponding SDF value.
Following ~\cite{parkDeepSDFLearningContinuous2019}, we then sample more aggressively near the zero-level-set of the $SDF_{roco}$ using a Gaussian function, $G(SDF_{roco}(\mathbf{x}),\sigma)$.
This strategy facilitates the network learning more detailed SDF near the object surface.

\subsection{Model Training and Loss Functions}
We train the network using two learning strategies: 1) self-supervised learning for $f_{ro}$ and $f_{co}$ in step (a) since there are no ground truth SDFs 
available during this stage and 
2) supervised learning for $f_{opt}$ in step (c) by using $SDF_{roco}$ as pseudo ground truth.

\begin{equation}
    \mathcal{L}_{step_a}=\mathcal{L}_{self}, \
    \mathcal{L}_{step_c}=\mathcal{L}_{super}
\end{equation}

\subsubsection{Self-supervised Learning}
The definition of self-supervised loss is mainly following UNSR~\cite{chenNeuralImplicitSurface2024}.
We additionally introduce a non-manifold regularizer to penalize query points that are not on the surface but are close to zero-level-set due to the constraints of $\mathcal{L}_{scc}$~\cite{yangStEikStabilizingOptimization2023}.

\begin{equation}
\begin{gathered}
     \mathcal{L}_{self}=\mathcal{L}_{sdf}+ \lambda_{scc} \mathcal{R}_{nonmfd}\mathcal{L}_{scc}  + \lambda_{adl}\mathcal{L}_{adl}, \\
    \mathcal{R}_{nonmfd} = \exp (-\alpha_{nonmfd}|f_{ro/co}(\mathbf{x})|), 
     \mathcal{L}_{sdf}= \left\|{\mathbf{x}}^{\prime}-\mathbf{\hat{x}}\right\|_2^2, \\
    L_{scc}= 1-\cos 
\left(\nabla f_{ro/co}({\mathbf{x}}),  ({\mathbf{x}^{\prime}} - {\mathbf{\hat{x}}})/{\left\|{\mathbf{x}}^{\prime}-\mathbf{\hat{x}}\right\|_2}\right)    
\end{gathered}
\end{equation}
where $\mathbf{\hat{x}} \in P$ is the nearest neighbor points of $\mathbf{x}$, $\nabla f_{ro/co}$ is the gradient of network $f_{ro}$ and $f_{co}$, $\mathcal{L}_{sdf}$ is an L2-distance loss to optimize the projected $\mathbf{x}^{\prime}$ in Eq.~\ref{EqProjection} reaching its nearest $\mathbf{\hat{x}}$, and $\mathcal{L}_{adl}$ is
the adversarial learning strategy in~\cite{chenNeuralImplicitSurface2024}.

\subsubsection{Supervised Learning}

The supervised loss is a L1-norm distance between the predicted SDF and $SDF_{roco}$ with a manifold regularizer~\cite{parkDeepSDFLearningContinuous2019,yangStEikStabilizingOptimization2023}.

\begin{equation}
\mathcal{L}_{super}=\left\|f_{opt}(\mathbf{x}) - SDF_{roco}\right\|_1 + \lambda_{mfd}\mathcal{R}_{mfd}, \mathcal{R}_{mfd} = \left\|{{f}_{opt}}(\mathbf{x})\right\|_1
\end{equation}
where $\mathcal{R}_{mfd}$ is a manifold regularizer to penalize query points away from the surface for smoothness.

\section{ Experiments and Results}

\subsection{Data Acquisition and Preparation}
Six computer-aided designed (CAD) vertebra models are used as phantoms from 3D printing,
including three typical thoracic vertebrae (T4, T8, T12) and three typical lumbar vertebrae (L1, L3, L5). 
These CAD models serve as ground truth for evaluation.
For the generalizability analysis, two different US transducers (UT1 and UT2) are adopted to collect two datasets using the row-column scan.
For one model scanned by one UT, each row-column scan corresponds to 1 shape and 2 scans. 
Two UTs obtain 12 shapes and 24 scans from 6 models. 
An average of 540$\pm$159 frames from UT1 and 686$\pm$244 frames from UT2 are collected. 
The electromagnetic (EM) positioning sensor is attached on the transducers 
for locating images in tracking space. 
The EM positioning sensor and UT are calibrated using the Levenberg-Marquardt algorithm~\cite{levenbergMethodSolutionCertain1944}.
More device parameters are listed in the supplementary material.

After the data acquisition, the 3D points in the region-of-interest (ROI) are segmented and transformed from the tracking space to a normalized 3D space, [-1,1].
Farthest Point Sampling (FPS)~\cite{qiPointNetDeepHierarchical2017} is applied to the row and column 3D point clouds for downsampling to speed up the training.
We randomly sample 20000 3D points from each segmented dataset using FPS and normalized them
to a 3D cube space
to acquire the raw point cloud.
Then, for each point in the point clouds, we sample 20 query points
following Gaussian distribution with mean 0 and standard deviation~\cite{chenNeuralImplicitSurface2024,NeuralPull}. 
The standard deviation is defined as the distance between $\mathbf{x}$ and its 
50\textsuperscript{th} nearest neighbor points.
Additionally, we randomly sample more query points within the same cube space to ensure the networks learn an SDF everywhere.

\begin{table}[tp]
  \caption{Performance comparison of our approach with the baseline on two datasets.}
  \renewcommand\arraystretch{0.9}  
  \centering
  \setlength{\tabcolsep}{2.5mm}{}
\begin{tabular}{ccccc}
\hline Methods & CD (mm)$\downarrow$ & HD (mm)$\downarrow$ & MAD (mm)$\downarrow$ & RMSE (mm)$\downarrow$ \\
\hline \multicolumn{5}{c}{ UT1 Scans } \\

\hline UNSR~\cite{chenNeuralImplicitSurface2024} (Row) & $2.16 \pm 0.16$ & $5.21 \pm 1.77$ & $1.84 \pm 0.14$ & $2.25 \pm 0.17$ \\
UNSR~\cite{chenNeuralImplicitSurface2024} (Col) & $2.11 \pm 0.18$ & $5.82 \pm 0.80$ & $1.78 \pm 0.17$ & $2.25 \pm 0.21$ \\
\textbf{RoCoSDF (Ours)} & $1.75 \pm 0.09$ & $4.08 \pm 0.74$ & $1.34 \pm 0.05$ & $1.70 \pm 0.03$ \\
\hline \multicolumn{5}{c}{ UT2 Scans } \\
\hline UNSR~\cite{chenNeuralImplicitSurface2024} (Row) & $2.40 \pm 0.62$ & $5.22 \pm 2.47$ & $1.97 \pm 0.67$ & $2.39 \pm 0.86$ \\
UNSR~\cite{chenNeuralImplicitSurface2024} (Col) & $2.54 \pm 0.63$ & $7.53 \pm 2.45$ & $2.25 \pm 0.69$ & $2.97 \pm 0.95$ \\
\textbf{RoCoSDF (Ours)} & $2.03 \pm 0.36$ & $4.87 \pm 2.80$ & $1.53 \pm 0.47$ & $1.92 \pm 0.74$ \\
\hline
\end{tabular}\label{Tab}

\end{table}

\begin{figure}[h]
	\centering
	\includegraphics[width=9cm]{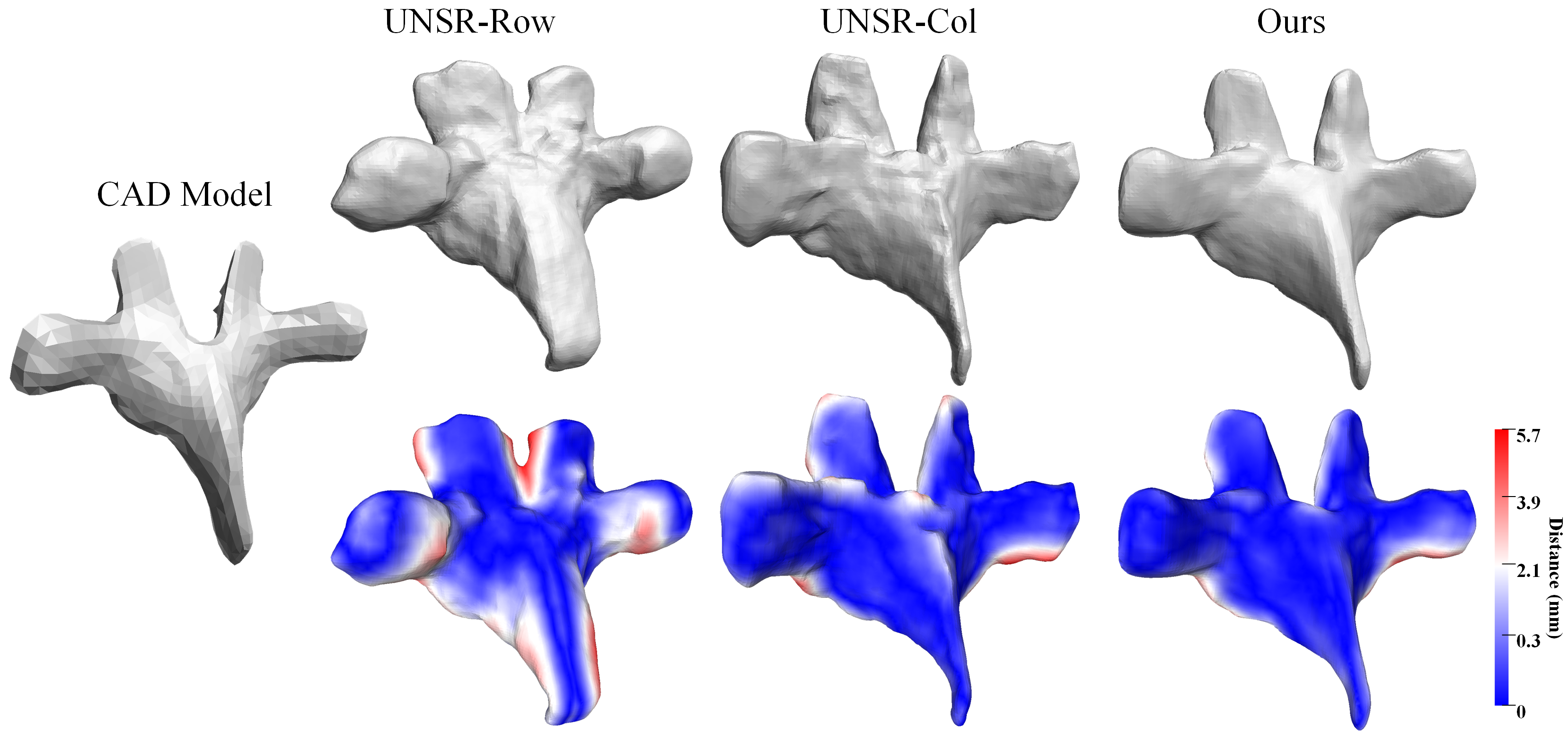}
	\caption{Visualization of thoracic vertebra T4. The top and bottom are the 3D meshes and colorized error maps from CAD model, UNSR (row-scan), UNSR (column-scan) and RoCoSDF (ours). 
         }
	\label{Fig_Result}
\end{figure}
\subsection{Evaluation Metrics}

Our approach is compared with the baseline method UNSR, the recent first method for the application of  
neural shape representation  to freehand 3D US imaging~\cite{chenNeuralImplicitSurface2024}. 
Four evaluation metrics are used to assess the reconstruction quality: Chamfer Distance (CD), Hausdorff Distance (HD), Mean Absolute Distance (MAD) and Root Mean Square Error (RMSE).
The distances are computed between the points randomly sampled from 
reconstructed mesh and the corresponding CAD models. 

We directly visually compare our approach with the traditional orientation-based reconstruction algorithms
due to the poor geometry appearance of traditional algorithms. 
The posed images from 
row-column scans are compounded using Pixel-Based Methods (PBM)~\cite{mohamedSurvey3DUltrasound2019,parkImprovingThreedimensionalAutomated2023}. 
The reconstructed volume is visualized through volume rendering.
We show the results in the supplementary material.

\subsection{Implementation Details}

Three neural networks, $f_{row}$, $f_{col}$ and $f_{opt}$, consist of 8 MLPs and 256 hidden channels with the skip connection at the fourth layer.
A Relu activation function is attached after each layer except for the last layer. 
We train $f_{row}$, $f_{col}$ and $f_{opt}$ for $1.0 \times 10^4$
iterations using the Adam optimizer.
The learning rate is set to 0.001 with a cosine decay schedule.
The batch size, $\lambda_{scc}$ and $\lambda_{adl}$ are set to 5000, 0.01 and 0.01,respectively.
The $\alpha_{nonmfd}$ and $\lambda_{mfd}$ and
are set to 100 and 0.6.
Our network is implemented using Pytorch and trained on a single NVIDIA RTX 3090
GPU with 24 GB memory. 
After the training, 
we set the mesh resolution to $256^3$ and the threshold of Marching Cube to 0  to extract the 
zero-level-set of surface boundaries using $SDF_{opt}$.

\subsection{Qualitative and Quantitative Results}

As listed in Table~\ref{Tab}, RoCoSDF achieves 
leading performance with UNSR across all four 
evaluation metrics on both two datasets.
Our approach demonstrates a 27\% and 24\% improvement over the UNSR row scan in UT1 scans in terms of MAD and RMSE, and 32\% and 35\% improvement over the UNSR column scan in UT2 scans in terms of MAD and RMSE.
More specifically, we observe that the quality of single-view reconstruction results from the UT2 scans dataset is inferior compared to those obtained from the UT1 scan dataset.
 This discrepancy in reconstruction performance can be attributed to the lower elevation resolution of the UT2 probe.
Nonetheless, our approach retains the capability to accurately recover the real structure from the row-column scan data, for example,
a substantial average decrease of 19\% and 35\% in RMSE than UNSR-row
 and UNSR-column.
Statistical significance is established with p-value < 0.01 against UNSR across all metrics.

We present a qualitative result of thoracic vertebra T4 in Fig.~\ref{Fig_Result} and more results in supplementary material. 
The reconstructed mesh  and the corresponding CAD model are superimposed
to visualize the error map with color coding.
The results from the single-view scanning by UNSR fail to reconstruct the accurate shape structure regardless of row-scan or column-scan,
with errors mainly distributed along the scanning direction. 
Compared to them, our method can extract latent shape information from each view to complementaryly recover accurate and smooth surfaces, as shown in the colorized error map at the bottom of Fig.~\ref{Fig_Result}.

\begin{figure}[tp]
	\centering
	\includegraphics[width=9cm]{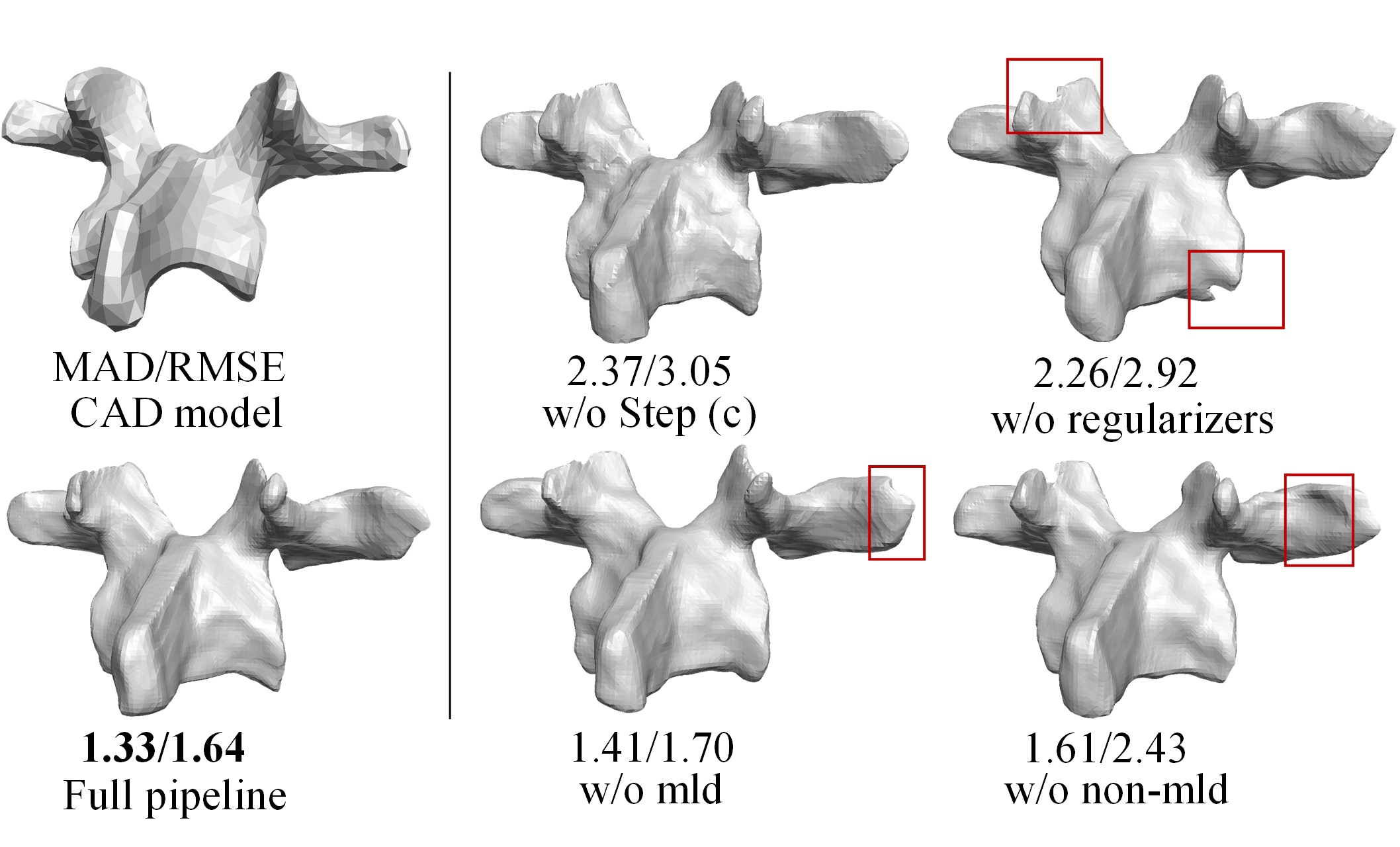}
	\caption{Ablation study on the effectiveness of step (c) and two regularizers.  
         }
	\label{Fig_AblationStudy}
\end{figure}

We qualitatively demonstrate an ablation study on lumbar vertebra L3 
to explore the effectiveness of step (c) and two regularizers.
As shown in Fig.~\ref{Fig_AblationStudy}, the generated meshes from direct CSG operation without step (c) yield obvious editing traces and noises.
The non-manifold regularization facilitates the network to learn a more compact shape and thin structures, while the manifold regularization refines the surface details.

\section{Conclusion}
We present RoCoSDF, a novel neural-SDF-based framework
for multi-view freehand 3D US shape reconstruction from row-column scanned data.
Our approach solves the challenges of view-dependent and pixel connectivity issues in multi-view 3D US by directly operating on 3D signed distance fields.
In addition, we design a coarse-to-fine optimization strategy to enhance
the shape appearance with surface regularizers, eliminating the need for additional ground truth shape supervision.
Evaluation results on two UTs demonstrate the high fidelity
and generalizability of our method.

The proposed RoCoSDF can effectively address the elevation thickness problem 
in freehand 3D US imaging as well as be easily extensible for robotic 3D US imaging.
This new design also holds promise 
to reconstruct precise shape from incomplete or noisy in-vivo structures by exploring other fusion strategies (such as union and interpolation) or denoising modules
with specific regularizations~\cite{marschnerConstructiveSolidGeometry2023,icml2020_2086}.




\begin{credits}
\subsubsection{\ackname} This work was supported by the Natural Science Foundation of China (No. 12074258).

\subsubsection{\discintname}
The authors have no competing interests to declare that are
relevant to the content of this article.
\end{credits}

%
%
%
%
\bibliographystyle{splncs04}
\bibliography{paper}

\begin{thebibliography}{10}
\providecommand{\url}[1]{\texttt{#1}}
\providecommand{\urlprefix}{URL }
\providecommand{\doi}[1]{https://doi.org/#1}

\bibitem{hennerspergerComputationalSonography2015a}
Hennersperger, C., Baust, M., Mateus, D., Navab, N.: Computational {{Sonography}}. In: Navab, N., Hornegger, J., Wells, W.M., Frangi, A. (eds.) Medical {{Image Computing}} and {{Computer-Assisted Intervention}} -- {{MICCAI}} 2015. pp. 459--466. Lecture {{Notes}} in {{Computer Science}}, {Springer International Publishing}, {Cham} (2015). \doi{10.1007/978-3-319-24571-3_55}

\bibitem{mohamedSurvey3DUltrasound2019}
Mohamed, F., Siang, C.V.: A {{Survey}} on {{3D Ultrasound Reconstruction Techniques}}. {IntechOpen} (Apr 2019). \doi{10.5772/intechopen.81628}

\bibitem{luoRecONOnlineLearning2023a}
Luo, M., Yang, X., Wang, H., Dou, H., Hu, X., Huang, Y., Ravikumar, N., Xu, S., Zhang, Y., Xiong, Y., Xue, W., Frangi, A.F., Ni, D., Sun, L.: {{RecON}}: {{Online}} learning for sensorless freehand {{3D}} ultrasound reconstruction. Medical Image Analysis  \textbf{87},  102810 (Jul 2023). \doi{10.1016/j.media.2023.102810}

\bibitem{luoMultiIMUOnlineSelfconsistency2023}
Luo, M., Yang, X., Yan, Z., Li, J., Zhang, Y., Chen, J., Hu, X., Qian, J., Cheng, J., Ni, D.: Multi-{{IMU}} with~{{Online Self-consistency}} for~{{Freehand 3D Ultrasound Reconstruction}}. In: Greenspan, H., Madabhushi, A., Mousavi, P., Salcudean, S., Duncan, J., {Syeda-Mahmood}, T., Taylor, R. (eds.) Medical {{Image Computing}} and {{Computer Assisted Intervention}} -- {{MICCAI}} 2023. pp. 342--351. Lecture {{Notes}} in {{Computer Science}}, {Springer Nature Switzerland}, {Cham} (2023). \doi{10.1007/978-3-031-43907-0_33}

\bibitem{huangAnatomicalPriorBased2022}
Huang, Q., Luo, H., Yang, C., Li, J., Deng, Q., Liu, P., Fu, M., Li, L., Li, X.: Anatomical prior based vertebra modelling for reappearance of human spines. Neurocomputing  \textbf{500},  750--760 (Aug 2022). \doi{10.1016/j.neucom.2022.05.033}

\bibitem{chenNeuralImplicitSurface2024}
Chen, H., Kumaralingam, L., Lou, E.H.M., Punithakumar, K., Li, J., Pham, T.T., Le, L.H., Zheng, R.: Neural {{Implicit Surface Reconstruction}} of {{Freehand 3D Ultrasound Volume}} with {{Geometric Constraints}} (Jan 2024). \doi{10.48550/arXiv.2401.05915}

\bibitem{parkImprovingThreedimensionalAutomated2023}
Park, C.K., Trumpour, T., Gyacskov, I., Bax, J.S., Tessier, D., Gardi, L., Ico, M., Fenster, A.: Improving three-dimensional automated breast ultrasound resolution with orthogonal images. In: Bottenus, N., Boehm, C. (eds.) Medical {{Imaging}} 2023: {{Ultrasonic Imaging}} and {{Tomography}}. p.~3. {SPIE}, {San Diego, United States} (Apr 2023). \doi{10.1117/12.2653141}

\bibitem{wrightFastFetalHead2023}
Wright, R., Gomez, A., Zimmer, V.A., Toussaint, N., Khanal, B., Matthew, J., Skelton, E., Kainz, B., Rueckert, D., Hajnal, J.V., Schnabel, J.A.: Fast fetal head compounding from multi-view {{3D}} ultrasound. Medical Image Analysis p. 102793 (Mar 2023). \doi{10.1016/j.media.2023.102793}

\bibitem{wysockiUltraNeRFNeuralRadiance2023}
Wysocki, M., Azampour, M.F., Eilers, C., Busam, B., Salehi, M., Navab, N.: Ultra-{{NeRF}}: {{Neural Radiance Fields}} for {{Ultrasound Imaging}} (Jan 2023). \doi{10.48550/arXiv.2301.10520}

\bibitem{gaitsUltrasoundVolumeReconstruction2024}
Gaits, F., Mellado, N., Basarab, A.: Ultrasound volume reconstruction from {{2D Freehand}} acquisitions using neural implicit representations. In: 21st {{IEEE International Symposium}} on {{Biomedical Imaging}} ({{ISBI}} 2024). p. {\`a} para\^{\i}tre. {IEEE Signal Processing Society and IEEE Engineering in Medicine and Biology Society}, {Ath{\`e}nes, Greece} (May 2024)

\bibitem{zhaEndoSurfNeuralSurface2023}
Zha, R., Cheng, X., Li, H., Harandi, M., Ge, Z.: {{EndoSurf}}: {{Neural Surface Reconstruction}} of~{{Deformable Tissues}} with~{{Stereo Endoscope Videos}}. In: Greenspan, H., Madabhushi, A., Mousavi, P., Salcudean, S., Duncan, J., {Syeda-Mahmood}, T., Taylor, R. (eds.) Medical {{Image Computing}} and {{Computer Assisted Intervention}} -- {{MICCAI}} 2023. pp. 13--23. Lecture {{Notes}} in {{Computer Science}}, {Springer Nature Switzerland}, {Cham} (2023). \doi{10.1007/978-3-031-43996-4_2}

\bibitem{batlleLightNeuSNeuralSurface2023}
Batlle, V.M., Montiel, J.M.M., Fua, P., Tard{\'o}s, J.D.: {{LightNeuS}}: {{Neural Surface Reconstruction}} in~{{Endoscopy Using Illumination Decline}}. In: Greenspan, H., Madabhushi, A., Mousavi, P., Salcudean, S., Duncan, J., {Syeda-Mahmood}, T., Taylor, R. (eds.) Medical {{Image Computing}} and {{Computer Assisted Intervention}} -- {{MICCAI}} 2023. pp. 502--512. Lecture {{Notes}} in {{Computer Science}}, {Springer Nature Switzerland}, {Cham} (2023). \doi{10.1007/978-3-031-43999-5_48}

\bibitem{parkDeepSDFLearningContinuous2019}
Park, J.J., Florence, P., Straub, J., Newcombe, R., Lovegrove, S.: {{DeepSDF}}: {{Learning Continuous Signed Distance Functions}} for {{Shape Representation}}. In: Proceedings of the {{IEEE}}/{{CVF Conference}} on {{Computer Vision}} and {{Pattern Recognition}}. pp. 165--174 (2019)

\bibitem{wang2021neus}
Wang, P., Liu, L., Liu, Y., Theobalt, C., Komura, T., Wang, W.: Neus: Learning neural implicit surfaces by volume rendering for multi-view reconstruction. NeurIPS  (2021)

\bibitem{amiranashviliLearningContinuousShape2024a}
Amiranashvili, T., L{\"u}dke, D., Li, H.B., Zachow, S., Menze, B.H.: Learning continuous shape priors from sparse data with neural implicit functions. Medical Image Analysis  \textbf{94},  103099 (May 2024). \doi{10.1016/j.media.2024.103099}

\bibitem{wiesnerGenerativeModelingLiving2024}
Wiesner, D., Suk, J., Dummer, S., Ne{\v c}asov{\'a}, T., Ulman, V., Svoboda, D., Wolterink, J.M.: Generative modeling of living cells with {{SO}}(3)-equivariant implicit neural representations. Medical Image Analysis  \textbf{91},  102991 (Jan 2024). \doi{10.1016/j.media.2023.102991}

\bibitem{NeuralPull}
Baorui, M., Zhizhong, H., Yu-Shen, L., Matthias, Z.: Neural-pull: Learning signed distance functions from point clouds by learning to pull space onto surfaces. In: International Conference on Machine Learning (ICML) (2021)

\bibitem{marschnerConstructiveSolidGeometry2023}
Marschner, Z., Sell{\'a}n, S., Liu, H.T.D., Jacobson, A.: Constructive {{Solid Geometry}} on {{Neural Signed Distance Fields}}. In: {{SIGGRAPH Asia}} 2023 {{Conference Papers}}. pp. 1--12. {ACM}, {Sydney NSW Australia} (Dec 2023). \doi{10.1145/3610548.3618170}

\bibitem{lorensenMarchingCubesHigh1987a}
Lorensen, W.E., Cline, H.E.: Marching cubes: {{A}} high resolution {{3D}} surface construction algorithm. ACM SIGGRAPH Computer Graphics  \textbf{21}(4),  163--169 (Aug 1987). \doi{10.1145/37402.37422}

\bibitem{sharmaCSGNetNeuralShape2018}
Sharma, G., Goyal, R., Liu, D., Kalogerakis, E., Maji, S.: {{CSGNet}}: {{Neural Shape Parser}} for {{Constructive Solid Geometry}}. In: Proceedings of the {{IEEE Conference}} on {{Computer Vision}} and {{Pattern Recognition}}. pp. 5515--5523 (2018)

\bibitem{yangStEikStabilizingOptimization2023}
Yang, H., Sun, Y., Sundaramoorthi, G., Yezzi, A.: {{StEik}}: {{Stabilizing}} the {{Optimization}} of {{Neural Signed Distance Functions}} and {{Finer Shape Representation}} (Nov 2023). \doi{10.48550/arXiv.2305.18414}

\bibitem{levenbergMethodSolutionCertain1944}
Levenberg, K.: A method for the solution of certain non-linear problems in least squares. Quarterly of Applied Mathematics  \textbf{2}(2),  164--168 (1944). \doi{10.1090/qam/10666}

\bibitem{qiPointNetDeepHierarchical2017}
Qi, C.R., Yi, L., Su, H., Guibas, L.J.: {{PointNet}}++: Deep hierarchical feature learning on point sets in a metric space. In: Proceedings of the 31st {{International Conference}} on {{Neural Information Processing Systems}}. pp. 5105--5114. {{NIPS}}'17, {Curran Associates Inc.}, {Red Hook, NY, USA} (Dec 2017)

\bibitem{icml2020_2086}
Gropp, A., Yariv, L., Haim, N., Atzmon, M., Lipman, Y.: Implicit geometric regularization for learning shapes. In: Proceedings of Machine Learning and Systems 2020, pp. 3569--3579 (2020)

\end{thebibliography}
\end{document}